\documentclass[runningheads]{svmult}

\usepackage{makeidx}   % allows index generation
\usepackage{graphicx}  % standard LaTeX graphics tool
\usepackage{subeqnar}  % subnumbers individual equations
\usepackage{multicol}  % used for the two-column index
\usepackage{physprbb}  % modified textarea for proceedings,
\makeindex             % used for the subject index
\begin{document}
\title*{350$\mu$m observations of local IRAS galaxies\protect\newline 
using SHARC-II}
\toctitle{350$\mu$m observations of local IRAS galaxies
\protect\newlineusing SHARC-II}
% allows explicit linebreak for the table of content
%
%
\titlerunning{SHARC-II observations of local IRAS galaxies}
% allows abbreviation of title, if the full title is too long
% to fit in the running head
%
\author{Colin Borys}
\authorrunning{Colin Borys}
% if there are more than two authors,
% please abbreviate author list for running head
%
%
\institute{California Institute of Technology, Pasadena CA 91125, USA}

\maketitle              % typesets the title of the contribution

\section{SEDs and redshifts of sub-mm bright galaxies }
The cosmologically significant population of dusty galaxies detected at sub-mm and mm wavelengths by SCUBA and MAMBO have played a central part in many of the discussions at this workshop.  Obtaining their redshifts is critical in understanding their role in galaxy formation and evolution, but this is a challenging task.   Nevertheless, Chapman et al. \cite{chapman} have made significant progress by obtaining spectroscopic redshifts of the subset of sub-mm galaxies detected in deep radio images that have optical counterparts. This exploits the well-known Far-Infrared/Radio correlation, and allows the high resolution radio maps to be used as a surrogate for finding the correct optical counterpart for which spectroscopy can then be attempted.

Because of the large investment of telescope time required to obtain and confirm the spectroscopic redshifts, and because the radio-detected sample consists only of the brighter sub-mm objects (only a few percent of the sub-mm source counts), some groups \cite{hughes,itzi,wiklind} have developed photometric techniques to estimate the redshifts of SCUBA/MAMBO detected galaxies.  These are also potentially biased because the Spectral Energy Distribution (SED) basically follows a modified black-body curve, which results in a fundamental degeneracy between redshift and dust temperature \cite{blain}.  However, by assuming that high-redshift galaxy SEDs resemble local ones, a library of templates derived from nearby galaxies can be used to essentially ``fix'' the temperature and break the degeneracy.  One problem with this approach is that the SEDs of local infrared-bright systems are poorly constrained for all but a few of brightest local sources.

One way to address this issue is to obtain photometry at different wavelengths. However, the atmosphere is opaque to FIR emission except for the narrow windows at relatively long wavelengths that SCUBA and MAMBO have already exploited.  Efficient observations at the shorter wavelength windows at 350 and 450$\mu$m have proved very difficult until now.

\section{SHARC-II observations of SLUGS galaxies}
The CSO recently commissioned the ``CCD-style'' SHARC-II camera \cite{sharc} optimized for use at 350$\mu$m. The combination of 384 sensitive detectors and a dish with low surface error allow observations previously impractical with other cameras operating at similar wavelengths.

Using SHARC-II, we have been imaging the sample of 106 IRAS selected galaxies in the highly successful SCUBA Local Universe Galaxy Survey (SLUGS) \cite{slugs1}.  These sources are lower luminosity analogues to the high-redshift SCUBA population.  With observations at 60, 100 (IRAS) and 850$\mu$m (SCUBA), the SEDs were fit and dust parameters estimated.  SCUBA was only able to detect 17 of these galaxies at 450$\mu$m, but from them it was recognized that the models required a component of cold dust to fit the the SED properly \cite{slugs2}.  These results highlight the need for more short-wavelength observations. To date, we have imaged 60 of the 106 SLUGS sources with SHARC-II and have and detected every one.  We are also conducting complementary observations of the Chapman et al. sample of SCUBA galaxies with spectroscopic redshifts.  Using this spec-z to break the T$/(1+z)$ degeneracy, the combination of SCUBA and SHARC-II observations will allow a study of the physical dust properties as a function of redshift.

\begin{figure}[b]
\begin{center}
\includegraphics[width=1.0\textwidth]{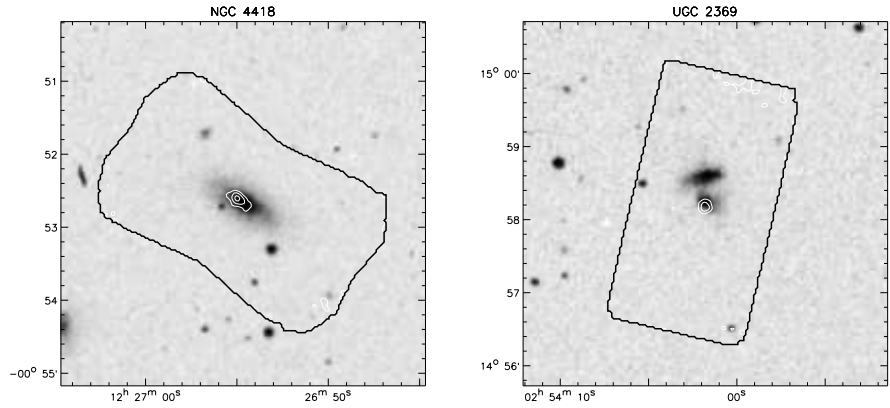}
\end{center}
\caption[]{Two IRAS galaxies imaged with SHARC-II.  Overlaid on DSS images are SHARC-II Signal-to-Noise contours at 4 and 6 sigma/beam.  The black outline denotes the area sampled by the SHARC-II array.}
\label{eps1}
\end{figure}

%INDEX%%%%%%%%%%%%%%%%%%%%%%%%%%%%%%%%%%%%%%%%%%%%%%%%%%%%%%%%%%%%%%%
% Please check with the editor of your book whether he plans to
% include a "mutual" subject index - if so, please code your entries
% in the standard syntax. For your own purposes you may print your
% "personal" index by using the following commands:
%
%\clearpage
%\addcontentsline{toc}{section}{Index}
%\flushbottom
%\printindex
%%%%%%%%%%%%%%%%%%%%%%%%%%%%%%%%%%%%%%%%%%%%%%%%%%%%%%%%%%%%%%%%%%%%%

\end{document}